\documentclass[12pt]{iopart}

\usepackage{iopams}  
\usepackage{epsfig}

\def\openone{\leavevmode\hbox{\small1 \normalsize \kern-.64em1}}

\begin{document}

\title[Interference contrast in multi-source few photon optics]{Interference contrast in multi-source few photon optics}

\author{Wies{\l}aw Laskowski$^1$, Marcin Wie{\'s}niak$^{1,2}$, Marek \.Zukowski$^1$, Mohamed Bourennane$^3$, Harald Weinfurter$^{4,5}$}

\address{${}^1$ Institute of Theoretical Physics and Astrophysics, University of Gda\'nsk, ul. Wita Stwosza 57, PL-80-952 Gda\'nsk, Poland} \address{${}^2$ Centre for Quantum Technologies, National University of Singapore, 3 Science Drive 2, Singapore 117543, Singapore} \address{${}^3$ Alba Nova Fysikum, University of Stockholm, S--106~91, Sweden} \address{${}^4$ Sektion Physik, Ludwig-Maximilians-Universit{\"a}t, D-80797 M{\"u}nchen, Germany} \address{${}^5$ Max-Planck-Institut f{\"u}r Quantenoptik, D-85748 Garching, Germany}

\ead{wieslaw.laskowski@univ.gda.pl}

\begin{abstract}
Many recent experiments employ several parametric down conversion (PDC) sources to get multiphoton interference.
Such interference has applications in quantum information. We study here how effects due to photon statistics, misalignment, and partial distinguishability of the PDC pairs originating from different sources may lower the interference contrast in the multiphoton experiments.
\end{abstract}


\section{Introduction}

Optical setups are an important implementation of various quantum information tasks, e.g. quantum teleportation \cite{T}, entanglement swapping \cite{ES} or quantum repeaters \cite{QR}. Most of the modern quantum optical experiments involving few photon interference use spontaneous parametric down-conversion (PDC) sources. It is a highly unlikely process in a non-linear crystal, in which a photon from the pumping beam changes into two entangled photons in the output modes. The quantum correlations are manifested in polarizations, momenta, and frequencies of the emitted pairs \cite{book, review}.

Recent years have brought a significant progress in realizing multisource PDC experiments allowing multiphoton interference. It was possible to experimentally verify five-photon entanglement \cite{FP}, and recently, six-photon interference was observed \cite{sixphoton}. However, these results required a considerable effort, putting in question the feasibility of realizing more complex systems. Therefore it is important to check if there are any fundamental obstacles for conducting future multiphoton interference experiments. Such obstacles might make some involved quantum information protocols impossible to  realize via purely photonic techniques.

Visibility (interferometric contrast) in multiphoton PDC experiments may be impaired by statistical properties of the emission process \cite{STAT}, misalignment and partial distinguishability of the entangled pairs. Therefore, a quantitative analysis of the influence of these factors is important, if we want to use non-classical properties of light in quantum communication. E.g., in order to show a conflict between quantum mechanics and classical description, the visibility should be greater than some critical value  $V^{crit}$. Different critical values might be needed to show that the state has entanglement of a required type. This report  investigates what is the impact of the specific properties of PDC on the observed multiphoton visibility, and what limitations follow from the necessity to breach these values.

We calculate the maximal possible visibilities under influence of these factors. Therefore, the description will be as simple as possible, as inclusion of other traits of the experiments would work toward lowering of the achievable visibility. To study the statistical properties we shall use the simplest description of type-II parametric down-conversion with just four modes of the radiation. We shall ignore completely the frequency-momentum structure of the states of the emitted signal-idler pairs. However, when studying effects due to distinguishability of such pairs originating from different emission acts, we shall ignore the statistical properties, and derive our results basing solely on the structure of the photon pair states. The other principal approximation that we shall make is neglecting in our description all sources of losses. Thus the problems of detection and collection efficiency will be ignored. 

The paper is organized in the following manner. 
In section II we study the decrease of visibility due to the production of the additional pairs of photons, which occurs in the strong pumping regime. This will be discussed for a Bell-type experiment and a setup designed to observe GHZ-correlations.  
Section III is divided in two parts. The first analyzes the problem of mode mismatch, which may occur if the interfering photons come from two different sources. 
In the second part we first give a simple description of the spectral properties of the effective two-photon output state of the spontaneous parametric down conversion process. This is followed by a consideration of the reduction of interferometric visibility due to the strong temporal correlations, which characterize a such a state. This affects the indistinguishability of the quantum processes associated with different emission acts, and therefore interference. The usual method of enforcing indistinguishability by employing suitable filters is studied quantitatively for a sequence of multiphoton experiments. We start with a single source experiment (two photon interference) and continue the analysis up to five sources (ten photon interference). 
In section IV we analyze the properties of the effective output state in PDC process. The basic result of this section is an observation that the reduction of the interferometric contrast due to (unwanted) multiple emissions, in some interesting cases, cannot be modeled by an additional white noise admixture to the observed interference pattern. The noise in some cases has a much more involved structure. We also give a new definition of an overall interferometric contrast, which seems to describe the situation much better than the standard one. It involves interferometric processes of various order.  
Section V contains conclusions and summary.

\section{The influence of statistics for high conversion efficiency}

To obtain multiphoton interference with several PDC sources one has to go to sufficiently high efficiencies of the down conversion process, which is usually described as a ``strong pumping regime'', because otherwise losses and detection inefficiency make the multiphoton coincidences prohibitively low. However, multiple emissions cause effects, which lower the visibility of the interferometric processes.

\subsection{Statistical properties of the PDC radiation}

\begin{figure}
\begin{center}
\includegraphics[width=0.35\textwidth]{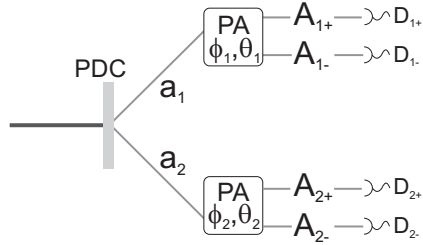}
\caption{The scheme for the polarization analysis of entangled pairs. PDC - parametric down--converting source, PA - polarization analyzer (the measurement settings are parametrized by $\theta_i$, $\phi_i$), a$_i$  - PDC output modes, A$_{i\pm}$ - analyzer output channels, D$_{i\pm}$ - detectors.}
\label{setup}\end{center}
\end{figure}

Assume that the PDC source emits pairs of polarization entangled photons into the spatial modes $a_1$ and $a_2$ (see Fig. \ref{setup}).
The most simplified form of the Hamiltonian for the PDC process can be put as
\begin{equation}
\mathcal{H} = i \chi (a_{1,H}^{\dagger} a_{2,H}^{\dagger} + a_{1,V}^{\dagger} a_{2,V}^{\dagger}) + h.c.
\end{equation}
Horizontally ($H$) and vertically ($V$) polarized photons occupy two spatial modes ($a_1,a_2$),
and $\chi$ is an effective coupling constant proportional, among others, to the pump power and the second-order nonlinearity of the crystal. After the interaction time the resulting photon state is given by $|\psi \rangle = e^{-iHt} |\Omega\rangle$, namely,
\begin{equation}
|\psi\rangle = \frac{1}{\cosh^4{K}} \sum_{n,m=0}^{\infty} \tanh^{n+m}{K} |n\rangle_{a_{1}}^H |m\rangle_{a_{1}}^V |n\rangle_{a_{2}}^H |m\rangle_{a_{2}}^V,
\label{state}
\end{equation}
where $|\Omega\rangle$ is the vacuum state, $K= \chi t$ and $|i\rangle_{x_y}$ denotes $i$-photon state in the spatial mode $x$ and polarization mode $y$, see e. g. \cite{SIMON-BOUWMEESTER}.
We shall now study the interference effects obtainable when one suitably overlaps radiation of several such sources. 

\subsection{A Bell-type experiment}

Let us first discuss  a standard Bell-type experiment involving just {\em one} source. One can pass the radiation given by eq. (\ref{state}) through polarization analyzers $PA_1$ and $PA_2$ (Fig. 1). 
The polarization measurements are performed by operators {\em corresponding} to qubit dichotomic
observables with eigenvectors 
$|\pm, \phi_i, \theta_i \rangle = \cos(\pm \frac{\pi}{4} + \theta_i)|H\rangle+ \sin(\pm \frac{\pi}{4} + \theta_i) e^{i \phi_i}|V\rangle$, ($i=1,2$) and eigenvalues $\pm 1$. The kets $|V\rangle$ and $|H\rangle$ represent {\em here} two orthogonal qubit states. One can write the annihilation operators $A_{i\pm}$ of the photons in the modes observed by the detectors in terms of the annihilation operators in the input modes of the polarization analyzers. This gives
\begin{eqnarray}
A_{1\pm}  = \cos(\pm \frac{\pi}{4} + \theta_1) a_{1,H} + \sin(\pm \frac{\pi}{4} + \theta_1) e^{i \phi_1} a_{1,V}  \nonumber \\
A_{2\pm}  = \cos(\pm \frac{\pi}{4} + \theta_2) a_{2,H} + \sin(\pm \frac{\pi}{4} + \theta_2) e^{i \phi_2} a_{2,V}.  \label{measurement}
\end{eqnarray}
We make the usual simplifying assumption that the probability $p_2(D_{1r_1}, D_{2r_2}| \phi_1,\phi_2,\theta_1,\theta_2)$ of  detectors $D_{1+}$ and $D_{2+}$  to click is proportional to
\begin{eqnarray}
&p(D_{1r_1}, D_{2r_2}| \phi_1,\phi_2,\theta_1,\theta_2) = \langle \psi | n_{A_{1r_1}} ~~ n_{A_{2r_2}} |\psi \rangle = \langle \psi | \mathcal{P}^{\dagger} \mathcal{P}  |\psi \rangle, &\label{prob-b}
\end{eqnarray}
where $\mathcal{P} = A_{2r_2}  A_{1r_1}$ and $n_X$ is the photon number operator in mode $X$. This approximation is justified for linear detectors, and allows us to get simple analytic formulas estimating the detection frequencies.
We obtain:
\begin{eqnarray}
&p_2 \sim p(D_{1r_1}, D_{2r_2}| \phi_1,\phi_2,\theta_1,\theta_2) & \nonumber \\
&= \frac{\cosh^2{K}}{-1 + 3 \cosh{2K}} \Big\{ \frac{1}{2}-  r_1r_2\frac{1}{2}\Big[\sin(2\theta_1)\sin(2\theta_2)& \nonumber \\
              &-  \cos(2\theta_1)\cos(2\theta_2) 
     \cos{(\phi_1 + \phi_2)}\Big] \Big\}+ \frac{\sinh^2{K}}{-1 + 3 \cosh{2K}}.&  \label{prob-true}
\end{eqnarray}
For details of the calculation see \ref{A-prob-true}. Note that for small values of $K$ the probabilities approach the ones  for an ideal maximally entangled two qubit state. 

\begin{figure}
\begin{center}
\includegraphics[width=0.4\textwidth]{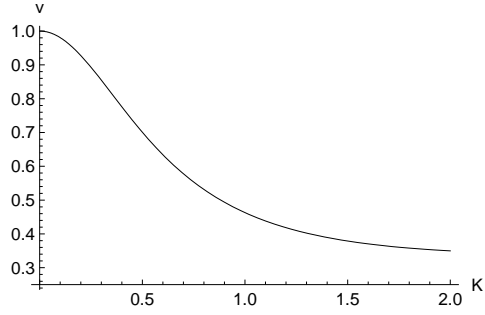}
\caption{The visibility of two photon interference in the experiment of Fig. 1 as the function of the PDC process efficiency parameter $K$.}
\label{pdc-2}
\end{center}
\end{figure}

As local detectors click at random, the formula for the ``two-photon'' interference visibility in this experiment can be put in a standard form
\begin{equation}
V_2 \equiv \frac{p^{max}_2 - p^{min}_2}{p^{max}_2 + p^{min}_2}
= \frac{1}{1 + 2\tanh^2 K}, \label{v-pdf}
\end{equation}
where the maximum (minimum) is taken over $\phi_a$, $\phi_b$, $\theta_a$, $\theta_b$. For small $K$, the visibility is close to 1 (see Fig. \ref{pdc-2}).
This case corresponds to a weak pumping field. In such a case a pump pulse usually produces nothing, and only sometimes a single pair. The  probability for other events is extremely low. If one increases the intensity of the pumping field, higher order (multifold) emissions start to play a role, and the visibility decreases.
For the unphysical case of infinite $K$, the visibility is equal to 1/3. A classical radiation of a thermal type gives maximal visibility of this value in two-detector experiments. All this recovers the results of \cite{KLYSHKO}. \\

\subsection{Multi source case: GHZ-states}

\begin{figure}
\begin{center}
\includegraphics[width=0.4\textwidth]{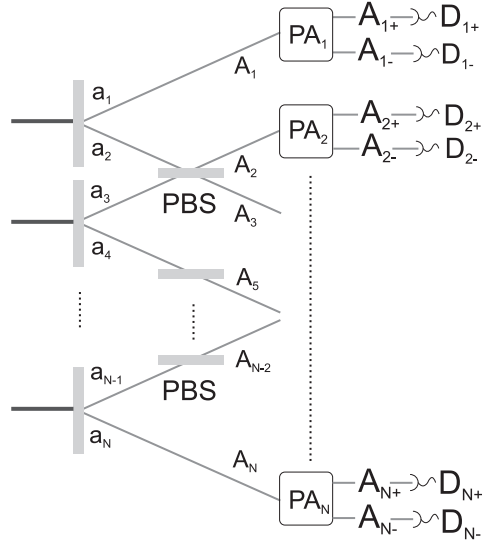}
\end{center}
\caption{An experiment with many PDC sources (compare fig. 1). The symbols a$_i$ (A$_i$) denote spatial modes  before (after) polarizing beam splitters (PBSs). A$_{i\pm}$ are  modes observed by detectors D$_{i\pm}$, which are behind polarization analyzers PA$_i$.}
\label{pdc-4}
\end{figure}

Let us now extend the above scheme by introducing additional sources (see Fig. \ref{pdc-4}).
Consider an $N$-photon experiment with $N/2$ sources. Modes $n$ and $(n+1)$ meet at a polarizing beam splitter (PBS). Again, the detectors are placed behind two-channel polarization analyzers.
A set-up of this kind was designed and used, for small pump power, to observe GHZ-type correlations \cite{a,b}.

The annihilation operators $A_{n\pm}$ of  the modes observed by detectors $D_{n\pm}$ can be written in terms of the annihilation operators 
in the input modes of the polarization analyzers in the following way:
\begin{equation}
A_{n\pm} = \frac{1}{\sqrt{2}}(A_n^H \pm e^{i \phi_n} A_n^V).
\label{pol-2}
\end{equation}
To simplify the discussion we assume that all $\theta_i$'s of the local settings are zero.
In turn, modes $A_n^H$ and $A_n^V$ ($n=1,...,N$) can be expressed in terms of the initial modes:
\begin{eqnarray}
&&A_1^H   = a_1^H; ~~~A_N^H  = a_N^H,\nonumber\\
&&A_{2n}^H   = a_{2n+1}^H;~~~ A_{2n+1}^H = a_{2n}^H, \label{pbs-ev}
\end{eqnarray}
where $n=1,...,N-1$ and $A_n^V = a_n^V$ for $n=1,...,N$. In these formulas we assume that H polarization is transmitted whereas V is reflected. 

\begin{figure}
\begin{center}
\includegraphics[width=0.4\textwidth]{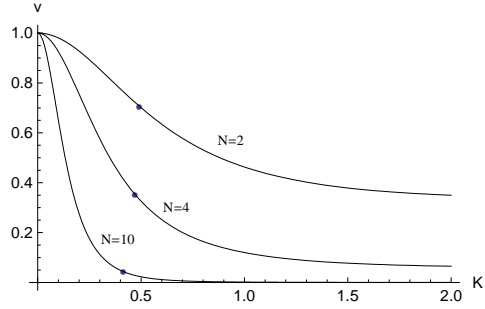}
\end{center}
\caption{Visibilities versus $K$ for multisource GHZ-type experiments. The points on the plots denote critical values beyond  which the visibility is not sufficient to violate standard Bell inequalities.}
\label{Graph1}
\end{figure}

Performing similar steps as in the previous case (see \ref{A-prob-true}) we get the probability of a joint detection in the form of 
\begin{eqnarray}
p_N &=& p(D_{1r_1}, ..., D_{Nr_N}| \phi_1, ...,\phi_N ) \nonumber \\
&\sim&  2 \tanh^N{K} \left(\prod_{i=1}^{N}r_i\cos\left(\sum_{j=1}^N \phi_j\right) + \tilde{p}_N\right),
\label{PROB-N}
\end{eqnarray}
where
\begin{equation}
\tilde{p}_{N}=\sum_{j=0}^{N}\frac{4^j}{(2j)!} \left[\prod_{k=0}^{j-1}(\frac{N^2}{4}-k^2)\right] \tanh^{2j}K.
\end{equation}
The first term in the sum is equal to 1, while the last one reads $2^{N-1}\tanh^{N}K$.
In particular, 
\begin{eqnarray}
\tilde{p}_2 &=& 1 + 2 \tanh^2{K}, \nonumber \\
\tilde{p}_4 &=& 1 + 8 \tanh^2{K} + 8 \tanh^4{K}, \nonumber\\
\tilde{p}_6 &=& 1 + 18 \tanh^2{K} + 48 \tanh^4{K} + 32 \tanh^6{K}, \nonumber\\
\tilde{p}_8 &=& 1 + 32 \tanh^2{K} + 160 \tanh^4{K} + 256 \tanh^6{K} \nonumber\\ &+& 128 \tanh^8{K},\nonumber\\
\tilde{p}_{10} &=& 1 + 50 \tanh^2{K} + 400 \tanh^4{K} + 1120 \tanh^6{K} \nonumber\\ &+& 1280\tanh^8{K} +  512\tanh^{10}{K}.
\end{eqnarray}
The visibility reads
\begin{equation}
\label{michelson}
V_N = \frac{p^{max}_N - p^{min}_N}{p^{max}_N + p^{min}_N} = \frac{1}{\tilde{p}_N}.
\end{equation}
In Fig. \ref{Graph1} we compare the visibilities $V_N$ as functions of $K$.
For the unphysical very high values of $K$ the visibilities asymptotically converge to the following value:
\begin{equation}
V_{N}(K\to \infty)=\frac{2}{(\sqrt{2}+1)^{N}+(\sqrt{2}-1)^{N}}.
\end{equation}

These values are related to modified Fibonacci numbers. Originally, they are elements of a sequence given by $F(0)=0,F(1)=1$ and $F(n)=F(n-1)+F(n-2)$ thereafter. In this case, the sequence satisfies $F'(0)=0,F'(1)=1,F'(n)=2F'(n-1)+F'(n-2)$. The solution reads 
\begin{equation}
F'(n)=\frac{(1+\sqrt{2})^n-(1-\sqrt{2})^n}{\sqrt{2}}.
\end{equation}
Thus
\begin{equation}
V_{N}(K\to \infty)=\frac{1}{F'(N/2)^2+(-1)^{N/2}}.
\end{equation}
For example,
\begin{eqnarray}
&V_2(K\to \infty)= 1/3 ~~~ V_4(K\to \infty)=1/17& \nonumber \\
&V_6(K\to \infty)=1/99   ~~~ V_8(K\to \infty)=1/577& \nonumber \\
&V_{10}(K\to \infty)=1/3363.& \nonumber
\end{eqnarray}

Note that for the analysis of these GHZ-like correlations we use only $N$-photon interference described by the probability formula (\ref{PROB-N}). Therefore, there is no ambiguity in the definition of $V_N$, as formula (\ref{PROB-N}) does not reveal any interference involving less than $N$ photons.  Simply, one can easily show that the marginals like   
$$
p(D_{1+},D_{2+}, ..., D_{N+})+p(D_{1-},D_{2+} ..., D_{N+})
  $$ do not depend on the phases.
Had there been lower order interference processes, one could face  difficulties in interpretation of the meaning of the parameter $V$. Thus far, these were fully resolved only in the two particle case \cite{jeager}.

We are most interested in the critical values of $K_{crit}$ such that for $K>K_{crit}$, the visibility in the experiment is not sufficient to violate the standard Bell inequalities \cite{WW,ZB,weinfurterzukowski}. These are given in Tab. \ref{tab1}. We also give critical values of $K$, which are necessary to reveal entanglement in noisy  GHZ states \cite{PITTENGER, SEPARABILITY}.
\begin{table}[!h]
\caption{Critical values of parameters corresponding to $V^{crit}$, which is necessary to exclude local realistic (upper rows) or separable descriptions (lower rows). Respective symbols denote: \newline
$1- p_{\Omega} = \tanh^2{K_{crit}}$ -- probability, that one, or more, pairs of photons are created at one source; 
\newline 
$p_1$ --  probability of a  single pair creation at one source; 
\newline
$(1-p_{\Omega })^{N/2}$ --  probability, that each source emits at least one pair of photons; 
\newline
$(p_1)^{N/2}$ -- probability, that each source emits exactly one pair of photons.}
\label{tab1}
\begin{center}
\begin{tabular}{ccccccc|c}
\hline 
$N$ & $V^{crit}$&  $K_{crit}$&$1-p_{\Omega }$& $p_1$ & $(1-p_{\Omega })^{N/2}$&$(p_1)^{N/2}$ \\ 
\hline 
2&$\frac{1}{\sqrt{2}}$&0.4911&0.2071&0.1642&0.207106&0.164213&local realism \\
&$\frac{1}{3}$ & $\infty$  & 1 & 0 & 1 & 0&entanglement \\ \hline
4&$\frac{1}{2\sqrt{2}}$&0.4697&0.1918&0.1550&0.036790&0.024030&local realism \\
&$\frac{1}{9}$ & 1.0613    &   0.6180  &  0.2360  &  0.381985  &  0.055726&entanglement \\\hline
6&$\frac{1}{4\sqrt{2}}$&0.4404&0.1714&0.1420&0.005033&0.002864&local realism \\
&$\frac{1}{33}$ & 0.9877   &  0.57218  &  0.2448  &  0.187258  &  0.014670&entanglement \\\hline
8&$\frac{1}{8\sqrt{2}}$&0.4232&0.1597&0.1342&0.000650&0.000324&local realism \\
 &$\frac{1}{129}$ & 0.9757   &  0.5643  &  0.2459  &  0.101400  &  0.003654&entanglement \\ \hline
10&$\frac{1}{16\sqrt{2}}$&0.4127&0.1527&0.1294&0.000083&0.000036&local realism \\
&$\frac{1}{513}$ & 0.9735    & 0.5629  &  0.2460   &  0.056493  &  0.000902&entanglement \\\hline
$\infty$& 0&0.3695&0.1250&0.10934&0&0&local realism\\
&0 & 0.9730    &  0.5625  &  0.2461  &  0  &  0&entanglement \\
\hline 
\end{tabular}
\end{center}
\end{table}

The values of $K_{crit}$ subtly decrease with $N$.
This means that for this family of experiments, the critical pump amplitude is always at almost similar level.
However, since the value of the $K$ for which we observe high enough interference is bounded by $K_{crit}$, the probability of an $n=N/2$-pair emission, proportional to  $ \left( \frac{\tanh^n{K}}{\cosh{K}} \right)^2$, is also bounded from above. Thus, since one cannot arbitrarily increase the pump power, with increasing $N$ one needs more and more time for the experiment. In Tab. \ref{tab1} we also present the critical values for the probability $p_1$ that only a single pair is created by a given source, the probability that each source emits exactly one pair (this a kind of an optimal event for the studied interference experiment), probability $1-p_{\Omega}$ that one has at least one pair emitted by a given source, and the probability that each source emits at least one pair of photons.   
In realistic experiments, the visibility also depends on other factors (see, e.g., section \ref{alignment}).  Thus,  parameter $K$ must be much lower not to additionally reduce the observed non-classical interference.

As shown above, statistics introduces a fundamental limit on the visibility in the multiphoton interferometric experiment, in which  we use pairs of entangled photons created in the parametric down conversion process. In order to obtain non-classical properties of down-converted photons the parameter $K_{crit}$ cannot be exceeded. It means that for experiments with many sources the  probability of a joint emission,  which supplies enough photons to observe a detection in each observation station, decreases exponentially with $n$ and the experiment must take much more time. To put it short, one cannot expect too much progress with multiphoton interference techniques similar to the one described here, without a significant increase in the collection and detection efficiency. High intensity pumping is counterproductive.

\section{Distinguishability problems}
Interference is perfect only if one has perfect indistinguishability of various quantum processes that lead to the detection events. In an experiment a mode mismatch might appear for  paths of propagation which are mixed (crossed) at a beamsplitter. This lowers the interference contrast.  Such a case is described in the first subsection. The other cause of degradation of interference, in the case of multi-source PDC experiments, is the strong energy-time correlation within a down-converted pairs. Via this property one could in principle distinguish photons originating from different sources. The second subsection presents a discussion of a  method of erasing such correlations with narrow spectral filters.  
\label{alignment}

\subsection{The problem of mode matching}

In experiments like the ones considered above one has to overlap radiation from different sources. This, causes alignment problems, which may reduce the visibility even further. We present here the simplest case. We shall see that misalignment introduces an independent factor reducing the visibility.

\begin{figure}[t]
\begin{center}
\includegraphics[width=\textwidth=0.75]{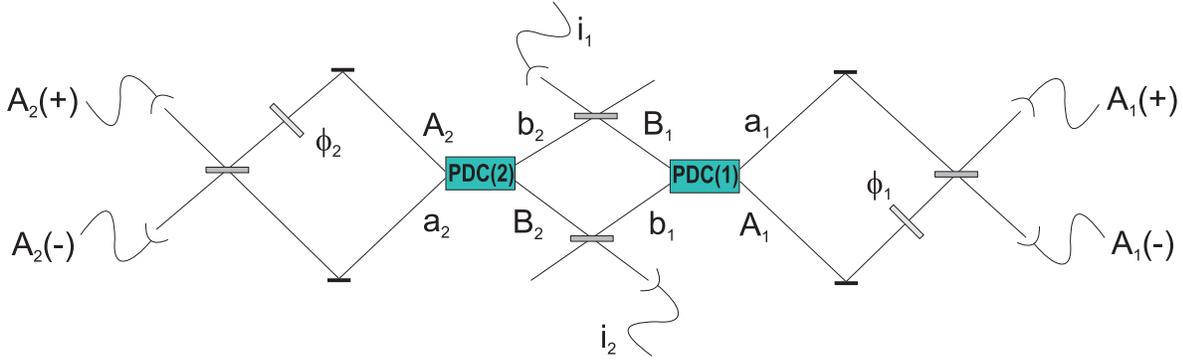}
\end{center}
\caption{Entanglement Swapping Scheme. Source PDC(1) emits photons either to spatial modes $a_1$ and $b_1$, or to $A_1$ and $B_1$. PDC(2)--to mode pairs $a_2,b_2$ or $A_2,B_2$. The swapping occurs when detectors $i_1$ and $i_2$ click simultaneously. Gray horizontal rectangles represent non-polarizing beam splitters, white diagonal are phase shifters, and black ones -- perfect mirrors.}
\label{star2}
\end{figure}

Consider the scheme presented in Figure \ref{star2}. It consists of two type-I PDC sources which produce photons of a fixed polarization in pairs of spatial modes $a_i$ and $b_i$, or $A_i$ and $B_i$ ($i=1,2$). Mode $b_1$ is crossed at a beam splitter with $B_2$, and similarly $b_2$ is crossed with $B_1$. Behind these beam splitters we place two detectors, labeled $i_1$ and $i_2$. In the ideal case, if both of them click simultaneously, the two outer photons are entangled. We now want to investigate how the two-photon interference attributable to this entanglement swapping is decreased by the fact that behind the beam splitters photons from modes $b_1$ and $B_1$ might be  partially distinguishable from those from $b_2$ and $B_2$, respectively. More precisely, a different signal arrives to a detector, depending whether the photon had been reflected or transmitted by the beam splitter. For example, the second source may produce photons with a slightly tilted polarization. Therefore, the annihilation operators related 	to the second crystal must be put as, e.g.,  $X'_2=\cos\alpha X_2+\sin\alpha X^\perp_2$, $(X=b,B)$. By $X^\perp_2$ we denote the radiation mode of source $2$ which is distinguishable behind the beamsplitter form the modes originating from the other source. 

Simple forms of the Hamiltonians describing the processes in the first and second source are given by:
\begin{eqnarray}
\mathcal{H}_{PDC(1)} &=& i \chi (a_1b_1 + A_1B_1) + h.c.,\nonumber \\
\mathcal{H}_{PDC(2)} &=& i \chi (a_2b'_2 + A_2B'_2) + h.c. 
\end{eqnarray}
In the Heisenberg picture the operators evolve as follows:
\begin{eqnarray}
a_2(T)&=&\cosh K  a_2 -\sinh K (\cos\alpha ~ b_2^\dagger+\sin\alpha (b_2^\perp)^\dagger), \nonumber\\
A_2(T)&=&\cosh K  A_2 -\sinh K (\cos\alpha ~B_2^\dagger+\sin\alpha (B_2^\perp)^\dagger),\nonumber\\
b_2(T)&=&\cosh K \cos\alpha (\cos\alpha~ b_2+\sin\alpha ~b_2^\perp)\nonumber\\
&-&\sinh K \cos\alpha a_2^\dagger,\nonumber\\
B_2(T)&=&\cosh K \cos\alpha (\cos\alpha ~B_2+\sin\alpha ~B_2^\perp)\nonumber\\
&-&\sinh K \cos\alpha A_2^\dagger,\nonumber\\
b_2^\perp(T)&=& \cosh K \sin\alpha (\cos \alpha ~b_2+\sin\alpha ~b_2^\perp)\nonumber\\
&-&\sinh K\sin\alpha ~  a_2^\dagger, \nonumber\\
B_2^\perp(T)&=& \cosh K \sin\alpha( \cos \alpha ~B_2+\sin\alpha ~B_2^\perp)\nonumber\\
&-&\sinh K\sin\alpha ~A_2^\dagger,
\end{eqnarray} 
and
\begin{equation}
X(T)=\cosh K X-\sinh K Y^\dagger,
\end{equation}
where $X=a_1,b_1,A_1,B_1$, and, respectively, $Y=b_1,a_1,B_1,A_1$. $X(T)$ denotes the annihilation operator in the $X$ mode after the interaction time $T$.

The probability of a detector click is proportional to the number of photons present
in the mode monitored by detector, i.e.
\begin{eqnarray}
&&p(A_1(+), A_2(+), i_1, i_2) \sim \langle: ~ n_{A_1(+)}(T) ~n_{A_2(+)}(T) \nonumber \\ 
&\times& (n_{i_1(+)}(T)+n_{i_1^{\perp}(+)}(T))~ (n_{i_2(+)}(T)+n_{i_2^{\perp}(+)}(T)) ~:\rangle,
\label{v-prob}
\end{eqnarray}
where $n_x(T) = x^{\dagger}(T) x(T)$ is the photon number operator in mode $x$ after the interaction time $T$. Note that we sum over in principle distinguishable detection event, thus we have here $n_i + n_i^{\perp}$ (recall that in our example $x$ and $x^{\perp}$ represent orthogonal polarizations.)

One can write the operators of the modes observed by the detectors in terms of the operators in the initial modes in the following way:
\begin{eqnarray}
&A_x(+) = \frac{1}{\sqrt{2}}(a_x + A_x e^{i\phi_x}) ~~(x=1,2),& \nonumber \\
&i_1    = \frac{1}{\sqrt{2}}(B_1 + b_2),  ~~~~~~ i_2    = \frac{1}{\sqrt{2}}(B_2 + b_1),& \nonumber \\
&i_1^{\perp}  = b_2^{\perp}, ~~~~~ i_2^{\perp}  = B_2^{\perp}.&  \\
\end{eqnarray}

\begin{figure}[!ht]
\begin{center}
\includegraphics[width=6cm]{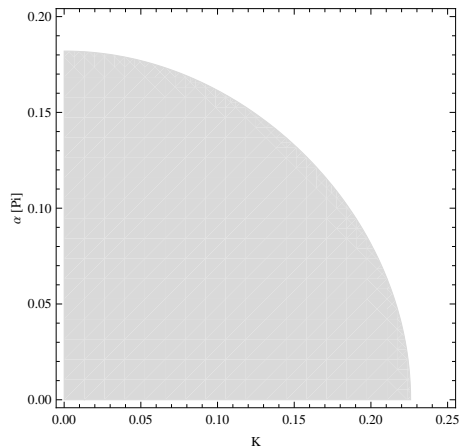}
\end{center}
\caption{The PDC efficiency parameters $K$ and the alignment parameters $\alpha$ for which $\tilde{V}>\frac{1}{\sqrt{2}}$ in the entanglement swapping scheme. Only in the shaded region the CHSH inequality can be violated.}
\label{independent}
\end{figure}

After some algebraic manipulations the final formula for the visibility of the detection rate (\ref{v-prob}) is given by:
\begin{equation}
\tilde{V}_4(K,\alpha) =  \cos^2{\alpha} ~ V_4(K).
\end{equation}
Figure \ref{independent} shows the  values of $K$ and $\alpha$ for which  the state created in the entanglement swapping scheme can violate a CHSH inequality \cite{CHSH}. The border values of $K$ correspond to the visibility equal to $1/\sqrt{2}$.

\subsection{Partial distinguishability of emission acts}

\label{temp}

If one considers the multisource scheme presented in Fig.\ref{pdc-4} one immediately sees that the tight frequency 
correlations of pairs of PDC photons originating from an individual act of emission may ruin our interference effects.
The frequencies of the members of a pair (photons labeled here as 1 and 2) sum up with great accuracy to the frequency of the pump, $\omega_p$,
\begin{equation}
\omega_1+\omega_2\approx\omega_p.
\end{equation}
This leads to strong correlations also in the time domain (such a phenomenon could be called ``an Einstein-Podolsky-Rosen'' effect''). Thus the pairs coming from different emission acts can be distinguishable in many ways - and interference suffers. 
This effect is independent of those studied in earlier Sections, where solely the influence of photon statistics was studied. It also requires a different calculational approach.
As it was suggested in \cite{ES} a proper use of filters and detection gates leads to a high interference contrast. Later on it was shown in \cite{ZZW, R} that the practical answer is to use pulsed pump and proper filters; this approach is used here.

We shall use the following simplified description of a two-photon state emitted by one of the sources that makes through the filters:
\begin{eqnarray}
\label{marek1}
|\psi\rangle&=&\int d\omega_1\int d\omega_2\int d\omega_0 g(\omega_0)f(\omega_1)f(\omega_2)\nonumber \\
&\times&(a^\dagger_H(\omega_1)a^\dagger_H(\omega_2)+a^\dagger_V(\omega_1)a^\dagger_V(\omega_2))\Delta(\omega_0-\omega_1-\omega_2)|\Omega\rangle,
\end{eqnarray}
where $a_P(\omega_x)$ is an annihilation operator describing a photon mode of frequency $\omega$ and polarization $P=H,V$. The index $x=1,2$ enumerates the directions, into which the photons are emitted. The symbol $f(\omega)$ denotes the spectral function of the filters, and $g(\omega_0)$ the spectral profile of the pump field. The operators satisfy the commutation relation $[a_P(\omega_x),a^\dagger_{P'}(\omega'_{x'})]=\delta_{PP'}\delta_{xx'}\delta(\omega-\omega')$. 
Such a simplified description of radiation (see e.g. Fearn and Loudon \cite{LandauFearn}) works quite well in the case when the directions of emission are well defined.
The function $\Delta$ represents the phase matching condition and will be replaced by  $\delta(\omega_0-\omega_1-\omega_2)$ in our considerations.

Next, if one puts the state into the form 
\begin{equation}
S^\dagger_{12}|\Omega\rangle=|\psi\rangle
\end{equation}
(where $S^\dagger_{12}$	denotes the operator in formula (\ref{marek1})), one sees that its generalization to one emission per source of Fig. 3 is given by $|\Psi\rangle = S^\dagger_{12}S^\dagger_{34}...S^\dagger_{N-1N}|\Omega\rangle$. This will be the starting point of our description. The probability that all $N$ detectors register a count, respectively at times $t_1,...,t_N$, is described by the usual formula for the $N$-fold detection frequency:
\begin{equation}
p(t_1,...,t_N)\propto\langle \Psi|E^{(-)}_{d_1}(t_1)...E^{(-)}_{d_N}(t_N)E^{(+)}_{d_N}(t_N)...E^{(+)}_{d_1}(t_1)|\Psi\rangle,
\end{equation} 
where the simplified effective field operators are given by
\begin{equation}
E^{(+)}_{d_k}(t_k)=\int d\omega_{d_k}e^{-i\omega_{d_k}t_k}a_{d_k}(\omega_{d_k}),  
\end{equation}
with indices $d_k$ denoting both the beam leading to the detector and the polarization, as each detector observes only radiation of a specified polarization.
The final annihilation operators $a_{d_k}$ are related to the initial ones by the usual algebra for the polarizing beam splitters and polarization analyzers.

Following Mollow \cite{Mollow}, $p(t_1,...t_N)$ can be written down using a square of the following amplitude:
\begin{equation}
\langle 0|E^{(-)}_{d_N}(t_N)...E^{(-)}_{d_1}(t_1)|\Psi\rangle.\nonumber
\end{equation}
This is due to the fact that $|\Psi\rangle$ is an $N$-photon state, therefore the action of $N$ annihilation operators leaves out only the vacuum component. Such an amplitude will be denoted by $\psi(t_1,...,t_N)$.

We assume that the detection stations measure elliptic polarizations, which can be represented by the following polarization mode transformation:
\begin{equation}
a_D(\omega)=\frac{1}{\sqrt{2}}(a_H(\omega)+e^{i\phi}a_V(\omega))
\end{equation}
(non polarization indices are dropped here). We assume that the polarizing beam splitters that mix the beams from two different sources  transmit the $H$ light and reflect the $V$ light.

An $N$-photon interference is possible only if: either all PDC photons are transmitted, or all down-converted photons are reflected by the mixing polarizing beam splitters (this effectively leads to a GHZ state, $(|H\rangle^{\otimes N}+|V\rangle^{\otimes N})/\sqrt{2}$). The interference depends on indistinguishability of the two processes. 
Thus we have an overall amplitude:
\begin{equation}
\psi(t_1,...,t_N)\propto\psi_R(t_1,...,t_N)+ e^{i\xi}\psi_T(t_1,...,t_N),
\end{equation}
where $e^{i\xi}$ represents the phases due to the measurement of the elliptic polarizations $\left(\xi=\sum_{i=1}^4\phi_{D_i}\right)$, and $\psi_R$ and $\psi_T$ are amplitudes with all photons reflected and transmitted, respectively.
One has
\begin{equation}
p(t_1,...,t_N)\propto |\psi_R|^2+|\psi_T|^2+2\textrm{Re}\left[e^{i\xi}\psi_R\psi_T\right].
\end{equation}
It the above formula we have assumed that $\psi$'s are real, which will be the case for amplitudes used below.

However, the detectors have a finite resolution time, which in comparison to the standard, in such experiments, pump pulse duration and the filter coherence time can be treated as infinitively long. Thus the overall detection probability behaves like
\begin{equation}
\int_{-\infty}^\infty dt_1...\int_{-\infty}^\infty dt_N p(t_1,...,t_N).\nonumber\\
\end{equation} 
Therefore, it is clear that the visibility 
is given by
\begin{eqnarray}
\label{vtempdef}
V_N&=&\Bigg| \int_{-\infty}^{\infty}dt_1...\int_{-\infty}^{\infty}dt_N \psi_R(t_1,t_2,t_3,...,t_{N-1},t_N)  \nonumber \\
&\times&  \psi_T(t_1,t_2,t_3,...,t_{N-1},t_N)\Bigg|.
\end{eqnarray}
Of course, above we have tacitly assumed normalization:
\begin{equation}
\int_{-\infty}^{\infty}dt_1...\int_{-\infty}^{\infty}dt_N |\psi_X(t_1,t_2,t_3,...,t_N)|^2=1
\end{equation}
for $X=R,T$. 

It is important to notice that
\begin{eqnarray}
\psi_R(t_1,t_2,...,t_N)=\psi(t_1,t_2)...\psi(t_{N-1},t_N)
\end{eqnarray}
and, similarly, due to the action of PBS's,
\begin{eqnarray}
\psi_T(t_1,t_2,...,t_N)=\psi(t_1,t_3)\psi(t_2,t_5)\psi(t_4,t_7)...\psi(t_{N-2},t_N),
\end{eqnarray}
where
\begin{eqnarray}
\psi(t,t')&\propto&\int d\omega'_0\int d\omega_1\int d\omega_2f(\omega_1)f(\omega_2)g(\omega'_0) \nonumber \\
&\times&\Delta(\omega'_0-\omega_1-\omega_2)e^{i\omega_1 t}e^{i\omega_2 t'}=\nonumber\\
&=&\langle\Omega|E^{(+)}_{1H}(t)E^{(+)}_{2H}(t')S^\dagger_{12}|\Omega\rangle
\end{eqnarray}
(the polarization was taken here as $H$ just for the sake of definiteness). 

If one assumes for simplicity that all relevant functions are Gaussian, that is 
\begin{equation}
\label{filter}
f(\omega)= \exp \left(-\left(\frac{\frac{\omega_0}{2}-\omega}{2\sigma_f}\right)^2\right)
\end{equation}
and
\begin{equation}
\label{pump}
g(\omega)=\exp\left(-\left(\frac{\omega_0-\omega}{2\sigma_0}\right)^2\right),
\end{equation}
one obtains that
\begin{eqnarray}
\label{timepair}
&& {\psi}(t_1,t_2) = \frac{1}{\sqrt{\Gamma}}\exp(\alpha_1(t_1^2+t_2^2)+\alpha_2t_1t_2),\\
&&\alpha_1 =-\frac{\sigma_0^2\sigma_f^2+\sigma_f^4}{\sigma_0^2+2\sigma_f^2}; ~~ \alpha_2 =2\frac{\sigma_f^4}{\sigma_0^2+2\sigma_f^2},\\
&&\Gamma =\frac{\pi}{2}\frac{\sqrt{\sigma^2_0+2\sigma_f^2}}{\sigma_0\sigma_f^2}.
\end{eqnarray}
Note, that  amplitudes of $\psi$, as promised, are real. 

We are now ready to calculate the visibility of $N$-photon interference effects. For all photons reflected by the polarizing beamsplitters, the amplitude $\psi_R$  takes the form
\begin{equation}
\label{temporal1}
{\psi}_R(t_1,...,t_N)=\frac{1}{\Gamma^{N/2}}\exp\left(\alpha_1\sum_{i=1}^Nt_i^2+\alpha_2\sum_{i=1}^{N/2}t_{2i-1}t_{2i}\right),
\end{equation}
whereas, for all photons transmitted we have
\begin{eqnarray}
\label{temporal2}
{\psi}_T(t_1,...,t_N)&=&\frac{1}{\Gamma^{N/2}}\exp\Big(\alpha_1\sum_{i=1}^Nt_i^2 \nonumber \\
&+&\alpha_2(t_1t_3+t_2t_5+t_4t_7+...+t_{N-2}t_N)\Big).
\end{eqnarray}
Please note that the visibility is a multiple integral of a Gaussian function. The argument of the exponent in ${\psi}_R(t_1,....t_N){\psi}_T(t_1,....t_N)$ is at most quadratic in all $t$'s. One can now extract the coefficients of a polynomial of $t_1$ and apply \begin{equation}
\label{fund}
\int_{-\infty}^\infty\exp(-\alpha x^2+\beta x+\gamma)dx=\sqrt{\frac{\pi}{\alpha}}\exp\left(\frac{\beta^2}{4\alpha}+\gamma\right),
\end{equation}
($\alpha>0$). Calculations of such a kind have to be done $N$ times (see Appendix B).
If one puts $f=\sigma_f/\sigma_0$, the formulas for visibilities read: 
\begin{eqnarray}
\label{Vtemp4}
V_4 &=& \frac{\sqrt{1+2f^2}}{1+f^2},\\
\label{Vtemp6}
V_6 &=& \frac{1+2f^2}{(1+f^2/2)(1+3f^2/2)},\\
\label{Vtemp8}
V_8 &=& \frac{2(1+2f^2)^{3/2}}{(1+f^2)(f^2+2+\sqrt{2})(f^2+2-\sqrt{2})},\\
\label{Vtemp10}
V_{10}&=& 16(1+2f^2)^2/(5(f^2+3+\sqrt{5})(f^2+3-\sqrt{5})\nonumber \\
&\times& (f^2+1+1/\sqrt{5})(f^2+1-1/\sqrt{5})),\\
\label{Vtemp12}
V_{12}&=& 16(1+2f^2)^{5/2}/((1+f^2)(2+f^2)(2+3f^2)\nonumber \\
&\times&(f^2+2(2+\sqrt{3}))(f^2+2(2-\sqrt{3}))).
\end{eqnarray}

\begin{figure}
\centering
\includegraphics[width=6cm]{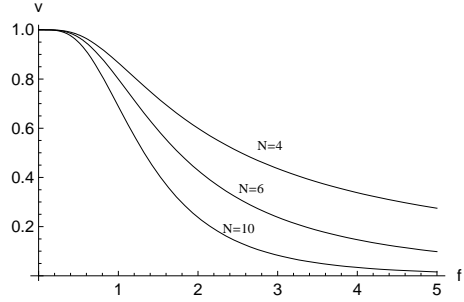}
\caption{The visibility of $N$-particle GHZ-type interference produced in the setup of Fig. \ref{pdc-4}. The curves are drawn for $N=4$ (the highest), 6, 10 (the lowest).}
\label{visib-f}
\end{figure}
For $f\ll 1$, that is narrow filters, $V^{temp}_N$ has the following tailor expansion:
\begin{equation}
V^{temp}_N\approx 1-\frac{f^4N}{8}+\frac{f^6N}4+...,
\end{equation}
while for broad filters, $f\gg 1$,
\begin{equation}
\label{largef}
V_N\approx \frac{2^{\frac{3N-2}4}}{N}f^{-\frac{N}2+1}.
\end{equation}
The last equation allows us to easily find a approximate critical value of $f$, above which local realism certainly cannot be falsified (with the use of WWW\.ZB inequalities \cite{WW, ZB, weinfurterzukowski}):
\begin{equation}
\label{fapprox}
f^{crit,approx}_N=\left(2^{1-\frac{5-N}{4}}N\right)^{\frac{2}{2-N}}.
\end{equation}
Table \ref{tab2} lists the critical visibility, required to violate the  WWW\.ZB inequalities, and the critical values of $f$ at which we have this visibility	obtained from (\ref{Vtemp4}-\ref{Vtemp12}) and (\ref{fapprox}).
\begin{table}
\caption{Values of $f^{crit}_N$ and $f^{crit,approx}_N$ for $N=4,6,8,10,12,\infty$.}
\label{tab2}
\begin{center}
\begin{tabular}{cccc}
\hline 
$N$ & $V^{temp}_{crit}$&  $f^{crit}_N$& $f^{crit,approx}_N$ \\ \hline 
4&$\frac{1}{2\sqrt{2}}$&3.806&4\\
6&$\frac{1}{4\sqrt{2}}$&3.592&3.884 \\
8&$\frac{1}{8\sqrt{2}}$&3.630&4\\
10&$\frac{1}{16\sqrt{2}}$&3.700&4.125 \\
12&$\frac{1}{32\sqrt{2}}$&3.737&4.237 \\
$\infty$& 0&?&$4\sqrt{2}$\\
\hline 
\end{tabular}
\end{center}
\end{table} 

As can be seen from Table \ref{tab2}, $f^{crit,approx}_N$ is an upper bound for $f^{crit}_N$. The approximation becomes less and less accurate with growing $N$. On the other hand, the sequence of $f^{crit}_N$ appears to be monotonously increasing for $N\geq 6$. Hence we can conjecture that $3.737<f^{crit}_\infty\leq 4\sqrt{2}$.

\section{Structure of Noise}

From (\ref{prob-true}) one can see that the effective, or ``apparent'', two qubit state produced in the two-photon experiment of Fig. 1 is a Werner state (a mixture of a maximally entangled state $|\phi^+\rangle=\frac{1}{\sqrt{2}}(|HH\rangle+|VV\rangle)$ and the ``white'' noise),
\begin{equation}
\rho_{eff} = V ~|\phi^+\rangle \langle \phi^+| + \frac{1-V}{4}, \label{NOISE}
\end{equation}
where $V$ is given by (\ref{v-pdf}) (we hope the usage of the same symbol for visibility and vertical polarization does not cause any trouble for a careful reader). 
By the effective state we shall mean an $N$-qubit state endowed with set of probabilities proportional to the one estimated in the actual experiment, and  which are afterward interpreted as ``$N$-photon'' coincidences. Interestingly, situation like the one of eq. (\ref{NOISE}) is not necessarily the case in experiments with more observers; the noise can be, in general, structured (or ``colored''). 

\begin{figure}
\centering
\includegraphics[width=6cm]{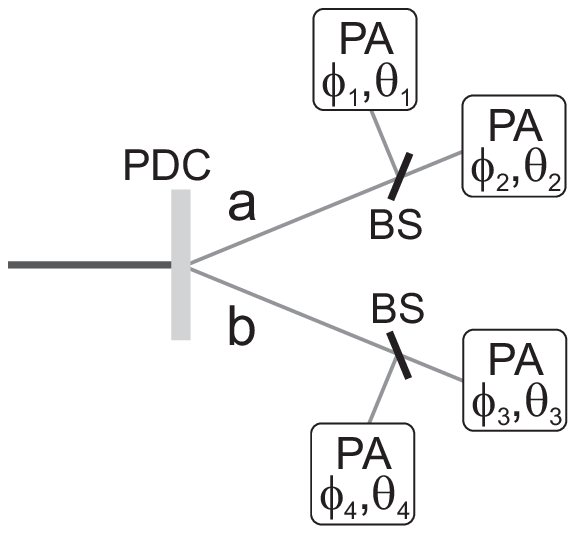}
\caption{The setup used in \cite{weinfurterzukowski} for generating correlations as of $|\psi_4\rangle$ state. PDC-nonlinear crystal, BS-beam splitter, PA-polarization analyzer.}
\label{PSI4}
\end{figure}

In this Section we aim to analyze the noise admixtures in the effective output state in the scheme of Ref. \cite{weinfurterzukowski}, presented in Fig. \ref{PSI4}. 
The setup consists of a single source of entangled photons, two non-polarizing beam-splitters (BS), and four polarization analyzers. If the source produces just two entangled pairs, in the state $\frac{1}{\sqrt{2}}(|HV\rangle+|VH\rangle)$, and one registers one photon in each of the detector, this collapses the initial state onto the so called $|\psi_4\rangle$ state:
\begin{eqnarray}
|\psi_4\rangle &=& \sqrt{1/3}(|HHVV\rangle +  |VVHH\rangle) \nonumber \\
&+& \sqrt{1/12}(|HVHV\rangle + |HVVH\rangle + |VHHV\rangle + |VHVH\rangle).
\label{psi}
\end{eqnarray}
The first two positions in the kets correspond to side $a$ in the figure, the other ones to side $b$.

However, this is an idealized situation, in which we do not take into account higher order emission processes. As we shall see these processes may influence experimenter's estimate which state was actually produced in the experiment. These corrections, due to the statistical properties of the PDC radiation are unavoidable, as the efficiency of the PDC process must be quite high to have a significant probability of having  of double pairs produced by a single pump pulse. This makes triple emissions also quite probable. Below we shall study the influence of  these statistical effects on the interpretation of the experimental data as far as the estimate of the final state is concerned. Other effects may also influence such an estimate, but they will be ignored.
  
The effective four qubit state observed in the experiment can be constructed via a tomographic method.
From the probabilities of registering clicks in detectors we construct the correlation function,
\begin{eqnarray}
&&E(\theta_1,\phi_1,...,\theta_4,\phi_4)= \sum_{r_1,r_2,r_3,r_4=\pm 1}r_1r_2r_3r_4 \nonumber \\ &&\times p(D_{1,r_1},D_{2,r_2},D_{3,r_3},D_{4,r_4}|\phi_1,\phi_2,\phi_3,\phi_4,\theta_1,\theta_2,\theta_3,\theta_4).
\label{corrf}
\end{eqnarray}
The probabilities here are now normalized, that is
\begin{eqnarray}
&\sum_{r_1,r_2,r_3,r_4=\pm 1} p(D_{1,r_1},D_{2,r_2},D_{3,r_3},D_{4,r_4}|\phi_1,\phi_2,\phi_3,\phi_4,\theta_1,\theta_2,\theta_3,\theta_4)=1.&
\end{eqnarray}
Note that the normalization factor is the probability that one observes a click at each of the four detection stations.
 
Subsequently, the values of the correlation function for specific sets of angles give us the elements of the correlation tensor,
\begin{equation}
T_{k_1,k_2,k_3,k_4}=E(\theta_{1,k_1},\phi_{1,k_1},\theta_{2,k_2},\phi_{2,k_2},\theta_{3,k_3},\phi_{3,k_3},\theta_{4,k_4},\phi_{4,k_4}),
\end{equation}
where
\begin{eqnarray}
&\theta_{i,k_i}=\left\{\begin{array}{ccc}0&\textrm{for}&k_i=1\\0&\textrm{for}&k_i=2\\\pi/4&\textrm{for}&k_i=3\end{array}\right.
,~~&\phi_{i,k_i}=\left\{\begin{array}{ccc}0&\textrm{for}&k_i=1\\\pi/2&\textrm{for}&k_i=2\\0&\textrm{for}&k_i=3\end{array}\right.
\end{eqnarray}
The three values of $k_i$ correspond, in the ``spin picture'' of a qubit, to three different complementary measurements associated with the Pauli matrices $\sigma_1=\sigma_x,\sigma_2=\sigma_y$ and $\sigma_3=\sigma_z$. We also need all marginal correlations, which are obtained from, for example,
\begin{eqnarray}
&&E(\theta_1,\phi_1,\theta_2,\phi_2,\theta_3,\phi_3)= \sum_{r_1,r_2,r_3,r_4=\pm 1}r_1 r_2 r_3 \nonumber \\
&&\times p(D_{1,r_1},D_{2,r_2},D_{3,r_3},D_{4,r_4}|\phi_1,\phi_2,\phi_3,\phi_4,\theta_1,\theta_2,\theta_3,\theta_4),
\label{Emarg}
\end{eqnarray}
That is we compute the marginals out of the observed four-fold coincidences. Please note that, the formula (\ref{Emarg}) is not the one obtained in the experiment with observation three-fold coincidences in detection stations 1,2 and 3, even if the detection efficiency is perfect. All reasonings are based on the assumption of registration of a four fold coincidence (which collapses the initial state).  The set of 256 numbers $\{T_{ijkl}\}_{i,j,k,l=0}^3$ allows one to reproduce the effective density matrix:
\begin{equation}
\rho_{eff}=\frac{1}{16}\sum_{i,j,k,l=0}^{3}T_{ijkl}\sigma_i^1\sigma_j^2\sigma_k^3\sigma_l^4,
\end{equation}
where $\sigma_0=\openone$. This procedure leads to
\begin{eqnarray}
\rho_{eff} &=& \frac{3 + 4 \cosh 2K + 5 \cosh 4K}{3 (3 - 4 \cosh 2K + 5 \cosh 4K)} |\psi_4 \rangle \langle \psi_4| \nonumber \\
&+& \frac{1 - 4 \cosh 2K + 5 \cosh 4K}{6 (3 - 4 \cosh 2K + 5 \cosh 4K)}
\sum_{i=1}^2 |\xi_i \rangle \langle \xi_i| \nonumber \\
&+&\frac{3 - 4 \cosh 2K + \cosh 4K}{6 (3 - 4 \cosh 2K + 5 \cosh 4K)} \sum_{i=3}^8 |\xi_i \rangle \langle \xi_i|,
\end{eqnarray}
where $|\xi_i\rangle$ are given by:
\begin{itemize}
\item \#1: $\frac{1}{2}(|HVVV\rangle + |VHVV\rangle + |VVHV\rangle + |VVVH\rangle)$,
\item \#2: $\frac{1}{\sqrt{2}}(|HHVV\rangle -|VVHH\rangle)$, 
\item \#3: $\frac{1}{2}(|HHHV\rangle + |HHVH\rangle + |HVHH\rangle + |VHHH\rangle)$, 
\item \#4: $|VVVV\rangle$, 
\item \#5: $\frac{1}{2}(-|HVVV\rangle - |VHVV\rangle + |VVHV\rangle + |VVVH\rangle)$,  
\item \#6: $\frac{1}{\sqrt{6}}(|HHVV\rangle - |HVHV\rangle - |HVVH\rangle - |VHHV\rangle - |VHVH\rangle + |VVHH\rangle)$, 
\item \#7: $\frac{1}{2}(-|HHHV\rangle - |HHVH\rangle + |HVHH\rangle + |VHHH\rangle)$, 
\item \#8: $|HHHH\rangle$. 
\end{itemize}

\begin{figure}
\centering
\includegraphics[width=8cm]{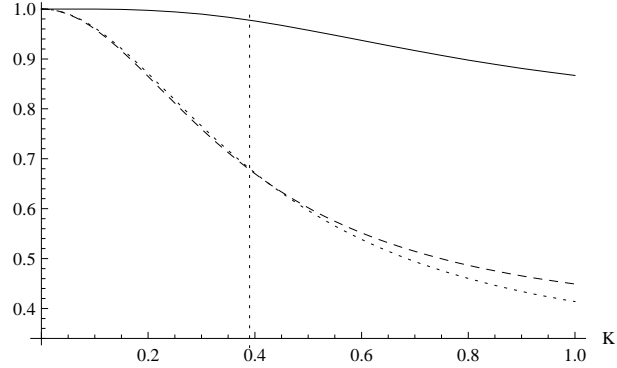}
\caption{The plot of the visibility (solid), the fidelity (dashed), and $V_{total}$ (dotted) of $\rho _{eff}$ as functions of $K$. The vertical line represents the critical visibility for the violation of the standard Bell inequality. Parameter $\epsilon$ is above 1 for all values of $K$.}
\label{vpf}
\end{figure}

The usual definition of visibility, the one used in earlier sections for GHZ correlations, does not reflect the quality of interference in such an experiment see Fig \ref{vpf}. Therefore we present also the fidelity of $\rho _{eff}$ with respect to the desired $|\psi_4\rangle$ state, $F=\langle\psi_4|\rho_{eff}|\psi_4\rangle$.  

A clearer picture of the interferometric properties of the state is given by the following function of purity
\begin{equation}
V_{total} = \sqrt{\frac{1}{2^N-1}({2^N}\textrm{Tr}\rho_{eff}^2 - 1)},
\end{equation}
where $N$ is the number of detecting stations (``qubits''). Note that $v_{total}$ varies between 1 and 0. The square root takes into account that correlations tensor components $T_{ijkl}$ enter $\textrm{Tr}\rho^2$ squared.
The parameters are presented in Fig. \ref{vpf}. Note that the formulas used for the visibility in the earlier section do not reflect here the real situation. It is clearly visible that $V_{total}$ is a much better parameter reflecting the loss of interference in this case.

The value of the parameter
\begin{equation}
\epsilon=\frac{\sum_{i,j,k,l}T_{ijkl}^2}{\max E},
\end{equation}
where, $\max E$ is the maximal value of the correlation function (\ref{corrf}) \cite{SEPARABILITY}, for the critical value $\epsilon=1$
gives one the threshold for having a sufficient condition for entanglement of the apparent state. 
Thus one could introduce yet another interferometric parameter of relevance. In the case studied here $\epsilon>1$ for all values of $K$. That is, the apparent state is always entangled.
Finally, $\max \sum T_{ijkl}^2$ with summations over pairs of orthogonal observable directions gives us the extent to which the given state may violate Bell inequalities \cite{mult}. These results will be presented elsewhere. 

Our discussion just a sketchy representation of the real situation. For real setups one must use a much more involved analysis, including a much more refined (and consistent) description of the detection process. Nevertheless, we think, the presented results signal specific problems that one might expect in real experiments. 

\section{Conclusions}

We have reviewed some difficulties that might be encountered in a typical few-photon PDC experiment. The possibility of creating more than the required number of pair of entangled photons in the parametric down-conversion, or  having some leftover frequency correlation between photons of the same pair, is unavoidable from the very fundamental point of view. 

There are some fundamental limitations for a successful multiphoton interference PDC experiment. The maximal value of the process efficiency parameter allowing to violate a standard Bell inequality decreases from $K(N=4)=0.4911$ to $K(N\rightarrow\infty)=0.3695$. This makes the production of at least one pair in each source exponentially improbable with a growing $N$. Hence, as a function of $N$, the whole experiment requires at least exponentially many runs, not only due to the growing number of measurements.

As pointed out in \cite{ES,ZZW,R}, the frequency correlations the photons can be removed by better defining their frequencies with narrow spectral filters. By the Heisenberg uncertainty principle, this causes the instants, at which the photons of a pair reach the detectors, to be sufficiently undefined to wash out the complementary temporal correlations. We have shown that one for pulse pumped sources cannot violate a standard Bell inequality for $\sigma_f / \sigma_0 >4\sqrt{2}$.

We  have also shown that in the case of some experiments, the noise introduced by unwanted additional emissions of photon pairs may have a quite complicated structure, which may affect the interpretation of the experimental results.

The description that we used was as simple as possible. We conjecture that refinements would lead to even more pronounced effects that those shown here. For specific experimental setups one must perform a more detailed analysis, involving a more realistic description of detectors response, and their inefficiency.         

\section*{Acknowledgments}

We thank Marcin Paw{\l}owski and Magdalena Stobi\'nska for discussions. We acknowledge the 
support by the DFG-Cluster of Excellence MAP, the DAAD/MNiSW exchange program,  Swedish Defense Material Administration (FMV) and Swedish Research Council (VR). The work is a part of the EU Projects
QAP (no. IST-015848) and SCALA. MZ was supported, at an early stage of this project, by Wenner-Gren Foundation.  
The work has been done at the {\em  National Quantum Information Centre of Gdansk}. 

\appendix

\section{Derivation of formula (\ref{prob-true})} \label{A-prob-true}

This appendix shows the way to calculate the formula (\ref{prob-true}) for $p(D_{1r_1}, D_{2r_2}| \phi_1,\phi_2,\theta_1,\theta_2)$.

Using (\ref{state}) and (\ref{measurement}) we obtain
\begin{eqnarray}
\mathcal{P}  |\psi \rangle &=& \frac{1}{\cosh^4{K}} \sum_{n,m=0}^{\infty} \tanh^{n+m}{K} \nonumber \\
&\times&\Big(\cos(\theta_1 + r_1\frac{\pi}{4})\cos(\theta_2 + r_2\frac{\pi}{4}) n \nonumber \\
&\times& |n-1\rangle_{a_{1}}^H |m\rangle_{a_{1}}^V |n-1\rangle_{a_{2}}^H |m\rangle_{a_{2}}^V \nonumber \\
&+&\cos(\theta_1 + r_1\frac{\pi}{4})\sin(\theta_2 + r_2\frac{\pi}{4}) \sqrt{nm} ~ e^{i\phi_2} \nonumber \\
&\times& |n-1\rangle_{a_{1}}^H |m\rangle_{a_{1}}^V |n\rangle_{a_{2}}^H |m-1\rangle_{a_{2}}^V \nonumber \\
&+&\sin(\theta_1 + r_1\frac{\pi}{4}) \cos(\theta_2 + r_2\frac{\pi}{4})\sqrt{nm}~e^{i\phi_1} \nonumber \\
&\times& |n\rangle_{a_{1}}^H |m-1\rangle_{a_{1}}^V |n-1\rangle_{a_{2}}^H |m\rangle_{a_{2}}^V \nonumber \\
&+&\sin(\theta_1 + r_1\frac{\pi}{4})\sin(\theta_2 + r_2\frac{\pi}{4})m~e^{i(\phi_1 + \phi_2)} \nonumber \\
&\times& |n\rangle_{a_{1}}^H |m-1\rangle_{a_{1}}^V |n\rangle_{a_{2}}^H |m-1\rangle_{a_{2}}^V\Big)
\end{eqnarray}
and the probability
\begin{eqnarray}
&&p(D_{1r_1}, D_{2r_2}| \phi_1,\phi_2,\theta_1,\theta_2) = \mathcal{N}^{-1} |\mathcal{P}  |\psi \rangle|^2 \nonumber \\
&=& \mathcal{N}^{-1} \frac{1}{\cosh^8{K}} \sum_{n,m=0}^{\infty} \sum_{n',m'=0}^{\infty} \tanh^{n+m+n'+m'}{K} \nonumber \\
&\times& \Big( \delta_{n-1,n'-1} \delta_{m,m'} \delta_{n-1,n'-1} \delta_{m,m'} \nonumber \\ 
&\times& \cos^2(\theta_1 + r_1\frac{\pi}{4})\cos^2(\theta_2 + r_2\frac{\pi}{4}) ~n n'\nonumber \\
&+& \delta_{n-1,n'-1} \delta_{m,m'} \delta_{n,n'} \delta_{m-1,m'-1} \nonumber \\ 
&\times&\cos^2(\theta_1 + r_1\frac{\pi}{4})\sin^2(\theta_2 + r_2\frac{\pi}{4}) \sqrt{n m n' m'}  \nonumber \\
&+& \delta_{n,n'} \delta_{m-1,m'-1} \delta_{n-1,n'-1} \delta_{m,m'} \nonumber \\ 
&\times&\sin^2(\theta_1 + r_1\frac{\pi}{4}) \cos^2(\theta_2 + r_2\frac{\pi}{4})\sqrt{n m n' m'} \nonumber \\
&+&\delta_{n,n'} \delta_{m-1,m'-1} \delta_{n,n'} \delta_{m-1,m'-1} \nonumber \\ 
&\times& \sin^2(\theta_1 + r_1\frac{\pi}{4}) \sin^2(\theta_2 + r_2\frac{\pi}{4}) m m'  \nonumber \\
&+& \frac{1}{4} \delta_{n-1,n'} \delta_{m,m'-1} \delta_{n-1,n'} \delta_{m,m'-1} \nonumber \\ 
&\times& \sin(2\theta_1 + r_1\frac{\pi}{2})\sin(2\theta_2 + r_2\frac{\pi}{2})n m'~ e^{i(\phi_1 + \phi_2)} \nonumber \\
&+&\frac{1}{4} \delta_{n'-1,n} \delta_{m',m-1} \delta_{n'-1,n} \delta_{m',m-1} \nonumber \\ 
&\times&  \sin(2\theta_1 + r_1\frac{\pi}{2})\sin(2\theta_2 + r_2\frac{\pi}{2}) n'm~ e^{-i(\phi_1 + \phi_2)}
\Big).
\end{eqnarray}
After the summation over $n'$ and $m'$ many terms vanish and we get the final formula for the probability:
\begin{eqnarray}
&&p_2 = p(D_{1r_1}, D_{2r_2}| \phi_1,\phi_2,\theta_1,\theta_2) \nonumber \\ 
&=& \mathcal{N}^{-1} \frac{1}{\cosh^8{K}} \sum_{n,m=0}^{\infty} \tanh^{2(n+m)}{K} \nonumber \\
&\times&\Big\{\cos^2(\theta_1 + r_1\frac{\pi}{4})\cos^2(\theta_2 + r_2\frac{\pi}{4}) ~n^2 \nonumber \\
&+& \sin^2(\theta_1 + r_1\frac{\pi}{4})\sin^2(\theta_2 + r_2\frac{\pi}{4})~m^2 \nonumber \\
&+& \cos^2(\theta_1 + r_1\frac{\pi}{4})\sin^2(\theta_2 + r_2\frac{\pi}{4})~nm \nonumber \\
&+& \sin^2(\theta_1 + r_1\frac{\pi}{4})\cos^2(\theta_2 + r_2\frac{\pi}{4})~nm \nonumber \\
&+&  \frac{1}{2} \sin(2\theta_1 + r_1\frac{\pi}{2})\sin(2\theta_2 + r_2\frac{\pi}{2}) ~n(m+1) 
                                                              \cos{(\phi_1 + \phi_2)}\Big\} \nonumber \\
&=& \mathcal{N}^{-1} \Big\{ \tanh^2{K} \Big[ \cos^2(\theta_1 + r_1\frac{\pi}{4})\cos^2(\theta_2 + r_2\frac{\pi}{4})
              \nonumber \\&+& \sin^2(\theta_1 + r_1\frac{\pi}{4})\sin^2(\theta_2 + r_2\frac{\pi}{4}) \nonumber \\
              &+& \frac{1}{2} \sin(2\theta_1 + r_1\frac{\pi}{2})\sin(2\theta_2 + r_2\frac{\pi}{2}) 
                                                              \cos{(\phi_1 + \phi_2)} \Big] \nonumber \\
              &+& \tanh^4{K} \Big[\cos^2(\theta_1 + r_1\frac{\pi}{4})\cos^2(\theta_2 + r_2\frac{\pi}{4}) 
              \nonumber \\&+& \sin^2(\theta_1 + r_1\frac{\pi}{4})\sin^2(\theta_2 + r_2\frac{\pi}{4}) \nonumber \\
              &+& \cos^2(\theta_1 + r_1\frac{\pi}{4})\sin^2(\theta_2 + r_2\frac{\pi}{4})
              \nonumber \\ &+& \sin^2(\theta_1 + r_1\frac{\pi}{4})\cos^2(\theta_2 + r_2\frac{\pi}{4})\Big] \Big\}                                                                            \label{prob-e}
\end{eqnarray}
Note that the terms in the bracket after ``$\tanh^4{K}$'' sum up to 1 and $\mathcal{N} = \frac{(-1 + 3 \cosh{2K})\tanh^2{K}}{\cosh^2{K}}$. The value of $\mathcal{N}$ we get using the normalization equation, namely  $\sum_{r_1,r_2=\pm1} p(D_{1r_1}, D_{2r_2}| \phi_1,\phi_2,\theta_1,\theta_2) =1$. Finally, using the above facts, and standard trigonometric relations, one can rewrite Eq. (\ref{prob-e}) to get the required final formula.

\section{Derivation of formula (\ref{Vtemp4})} \label{A-PHI}

By eq. (\ref{vtempdef}), the integral of our interest reads:
\begin{eqnarray}
&V_4=\frac{1}{\Gamma^2}\int_{-\infty}^\infty dt_1\int_{-\infty}^\infty dt_2\int_{-\infty}^\infty dt_3\int_{-\infty}^\infty dt_4 &\nonumber \\
&\times\exp\Big[-\frac{2\sigma_f^2} {\sigma_0^2+2\sigma_f^2} \Big((\sigma_0^2+\sigma_f^2)\sum_{i=1}^4t_i^2
-\sigma_f^2(t_1+t_4)(t_2+t_3)\Big)\Big].&
\end{eqnarray} 
The argument of the exponent can be put as
\begin{eqnarray}
f_1(t_1)&=&-2\sigma_f^2\frac{\sigma_0^2+\sigma_f^2}{\sigma_0^2+2\sigma_f^2}t_1^2
+\frac{2(t_2+t_3)\sigma_f^4}{\sigma_0^2+2\sigma_f^2}t_1 \nonumber \\
&-&2\sigma_f^2\frac{(\sigma_0^2+\sigma_f^2)\sum_{i=2}^4t_i^2-\sigma_f^2t_4(t_2+t_3)}{\sigma_0^2+2\sigma_f^2}.
\end{eqnarray}
We have singled out the part of the expression which depends on $t_1$. After integration over $t_1$, which employs
(\ref{fund}), we obtain
\begin{eqnarray}
V_4&=&\frac{\sqrt{\pi}}{\Gamma^2\sqrt{2\sigma_f^2\frac{\sigma_0^2+\sigma_f^2}{\sigma_0^2+2\sigma_f^2}}}	
\int_{-\infty}^{\infty}dt_2\int_{-\infty}^{\infty}dt_3\int_{-\infty}^{\infty}dt_4 \nonumber \\ 
&\times& \exp \Big[-\frac{2\sigma_f^2}{(\sigma_0^2+2\sigma_f^2)}
\Big((\sigma_0^2+\sigma_f^2)\sum_{i=2}^4t_i^2 \nonumber \\
&-& \sigma_f^2t_4(t_2+t_3) +\frac{(t_2+t_3)^2\sigma_f^4}{\sigma_0^2+\sigma_f^2}\Big)\Big].\label{EQUATION-F}
\end{eqnarray}
Again, we define the following quadratic form $f_2(t_4)$, in which $t_4$ is singled out:
\begin{eqnarray}
f_2(t_4)&=&-2\sigma_f^2\frac{\sigma_0^2+\sigma_f^2}{\sigma_0^2+2\sigma_f^2}t_4^2
+2\frac{(t_2+t_3)\sigma_f^4}{\sigma_0^2+\sigma_f^2}t_4 \nonumber \\
&-&\sigma_f^2\frac{-2(t_2+t_3)^2\sigma_f^4+(t_2^2+t_3^2)(\sigma_0^2+\sigma_f^2)^2}{2(\sigma_0^4+3\sigma_0^2\sigma_f^2+2\sigma_f^4)}.
\end{eqnarray}
and integrate in (\ref{EQUATION-F}) over $t_4$. This brings our integral to the form
\begin{eqnarray}
V_4&=&\frac{\sqrt{\pi}}{\Gamma^2(2\sigma_f^2\frac{\sigma_0^2+\sigma_f^2}{\sigma_0^2+2\sigma_f^2})}
\int_{-\infty}^{\infty}dt_2\int_{-\infty}^{\infty}dt_3\exp\Big[-\frac{2\sigma_f^2(t_2+t_3)^2}{(\sigma_0^2+\sigma_f^2)(\sigma_0^2+2\sigma_f^2)}\nonumber\\
&-&\frac{2\sigma_f^2(\sigma_f^4(t_2+t_3)^2+(\sigma_0^2+\sigma_f^2)^2(t_2^2+t_3^2)}{\sigma_0^4+3\sigma_0^2\sigma_f^2+2\sigma_f^4}\Big].
\end{eqnarray}
We put the function in the exponent into the following form
\begin{eqnarray}
f_3(t_3)&=&-\frac{2(\sigma_0^4+2\sigma_0^2\sigma_0^2+3\sigma_f^4)}{\sigma_0^4+3\sigma_0^2\sigma_f^2+2\sigma_f^4	}\sigma_f^2t_3^2\nonumber \\
& - &\frac{8\sigma_f^6t_2}{\sigma_0^4+3\sigma_0^2\sigma_f^2+2\sigma_f^4	}t_3\nonumber \\
& - &\frac{2(\sigma_0^4+2\sigma_0^2\sigma_0^2+3\sigma_f^4)}{\sigma_0^4+3\sigma_0^2\sigma_f^2+2\sigma_f^4}\sigma_f^2t_2^2.	
\end{eqnarray} 
Therefore, the integral reduces to:
\begin{eqnarray}
V_4&=&\frac{1}{\Gamma^2\sigma_f^3}\sqrt{\frac{\pi^3(\sigma_0^2+2\sigma_f^2)^3}{(\sigma_0^2+\sigma_f^2)(\sigma_0^4+2\sigma_0^2\sigma_f^2+3\sigma_f^4)}} \nonumber \\
&\times&\int_{-\infty}^{\infty}dt_2\exp\left(-2\sigma_f^2\frac{\sigma_0^2+\sigma_f^2}{\sigma_0^2+2\sigma_f^2}t_2^2\right),
\end{eqnarray}
and after purely algebraic simplifications gives (\ref{Vtemp4}).

\end{document}